\documentclass[9pt,twocolumn,twoside]{opticajnl}
\journal{opticajournal} 

\setboolean{shortarticle}{true}

\usepackage{lineno}

\title{Analog phase-sensitive time-reversal of optically-carried radiofrequency signals}

\author[1]{Thomas Llauze}
\author[1,*]{Anne Louchet-Chauvet}

\affil[1]{ESPCI Paris, Universit\'e PSL, CNRS, Institut Langevin, 75005 Paris, France}

\affil[*]{anne.louchet-chauvet@espci.fr}

\begin{abstract}
Achieving low-latency time-reversal of broadband radiofrequency signals is crucial for reliable  communications in dynamic, uncontrolled  environments. However, existing approaches are either digitally assisted --making broadband extension challenging-- or limited to amplitude modulation. In this work, we report the very first experimental realization of a fully analog, phase-preserving time-reversal architecture for optically-carried radiofrequency signals. The method exploits the exceptional coherence properties of rare-earth ion-doped materials, and leverages the well-established photon echo mechanism, widely used in quantum technologies. While our demonstration is conducted with a modest bandwidth, we identify the fundamental cause of this limitation and propose solutions for future scalability.
\end{abstract}

\setboolean{displaycopyright}{false} 

\begin{document}

\maketitle

\section{Introduction}
Time-reversal is an efficient way to focus waves in a complex environment~\cite{fink1992}. It was initially proposed for acoustic waves and resulted in very diverse applications, ranging from medical imaging and therapy~\cite{thomas1996lithotripsy}, to geoscience~\cite{micolau2003dort} and underwater communications~\cite{higley2005synthetic}.
More recently, time-reversal has been extended to radiofrequency waves~\cite{yavuz2009ultrawideband} and successfully demonstrated in pratical demonstrations in a reverberating chamber~\cite{lerosey2004time,lerosey2006time}.

In dynamic environments such as urban landscape, a fast time-reversal operator is essential for real-time adaptation to the medium. However, digital approaches are constrained by the limitations of digital-to-analog converters (DAC). Indeed, despite ongoing advancements in bandwidth, processing speed and memory depth, these figures of merit cannot be simultaneously optimized, and some compromise must be found. Specifically, in the case of broadband RF communication in a non-stationary, uncontrolled medium including moving objects at speeds up to $20$~m/s, achieving sub-ms latency is crucial, yet remains challenging with multi-GHz-bandwidth DACs.

One can by-pass the digitization step altogether by resorting to purely analog solutions. Among these, microwave photonics-based systems~\cite{yao2009microwave} stand out, leveraging the high bandwidth and low losses of photonic components. The analog time-reversal of an optically-carried signal was first obtained using a dispersive optical medium, namely  fibers~\cite{coppinger1999}, Bragg gratings~\cite{zhang2015}, or spectral gratings in an inhomogeneously broadened atomic absorption line~\cite{linget2013}. Their analog nature ensures minimal latency (a few $\mu$s at most) and the duration of the signals that can be processed, related to the dispersive power of the optical medium, can reach values as high as several $\mu$s. Nevertheless, by construction, these methods are only compatible with monochromatic, amplitude-modulated RF signals, since they require a univocal time-to-frequency correspondence in the optical domain.

In this work, we explore the capabilities of an original analog time-reversal protocol based on coherent light-matter interaction in a rare-earth ion-doped crystal. We show that unlike the previous demonstrations, this approach enables the retrieval of the radiofrequency field in amplitude and phase. 

\section{Analog signal processing architectures derived from photon echo}
\label{sec:symmetrical}
\subsection{Principle and examples}
Three-pulse photon echo (3PPE) is a four-wave mixing process operating in the temporal domain. When three time-separated fields are shined onto an inhomogeneously broadened medium, the latter radiates a delayed coherent response (the "echo") whose time-dependent field (to the lowest order) reads as:
\begin{equation}
\mathcal{E}_{echo}(t)=\mathcal{E}_1^*(-t) \otimes \mathcal{E}_2(t)  \otimes \mathcal{E}_3(t)
\label{eq:the_equation}
\end{equation}
A broad variety of microwave photonics-related signal processing architectures have been proposed based on this photon echo mechanism, including data bit-rate conversion~\cite{mossberg1982}, spectral analysis~\cite{crozatier2004tf, berger2016rf}, arbitrary signal generation~\cite{damon2010awg}, correlative processor~\cite{barber2017spatial}, and time-reversal of amplitude-modulated optical signals~\cite{linget2013,wang1995}. They always function as follows: the input, optically-carried information serves as the first, second or third field. The remaining two fields are shaped by the user, enabling the spontaneous generation of the echo carrying the processed output. Such architectures are by essence immune to computation delay. Their spectral bandwidth and resolution are ultimately limited by the width of the inhomogeneous atomic absorption line and by the homogeneous linewidth of the selected optical transition, respectively. They can reach highly attractive values (several tens or hundreds of GHz for the former, a few kHz for the latter), combined with a processing time of a few $\mu$s, ensuring competitiveness with digital approaches.

\subsection{Time-reversal scheme}
We propose a similar approach to achieve time-reversal of optically-carried RF signals. This method directly leverages the explicit time-reversal operation in equation~\ref{eq:the_equation}. The input signal serves as the first excitation field $\mathcal{E}_1(t)$, while the second and third excitation fields $\mathcal{E}_2(t)$ and $\mathcal{E}_3(t)$ (referred to as "control fields") are designed as two successive linear frequency chirps with opposite slopes $\mathcal{E}_2(t)=e^{i\pi r t^2}$ and  $\mathcal{E}_3(t)=e^{-i\pi r t^2}$. This way, the impulse response of the time-reversal filter, given by the  convolution of the two control fields, produces a short pulse $\mathcal{E}_2(t)  \otimes \mathcal{E}_3(t) \simeq\delta(t)$, leading to :
\begin{equation}
\mathcal{E}_{echo}(t) \simeq \mathcal{E}_1^*(-t)
\label{eq:E1}
\end{equation}
In practice, the finite duration of the control fields leads to a temporal broadening of the impulse response. For example, when the control fields have a rectangular temporal shape with  duration $T$, the impulse response reads as: $\mathcal{E}_2(t)  \otimes \mathcal{E}_3(t) \simeq \frac{\sin (\pi r T t)}{\pi r t} $.


This scheme closely resembles previous photon-echo-based time-reversal protocols, with the key distinction that the input signal is carried by the first pulse, rather than the second~\cite{wang1995optical,mossberg1982} or third pulse~\cite{linget2013}. Though seemingly minor, this has important consequences.
First, our approach does not require a chirped signal carrier, making it compatible with arbitrary fields. In contrast, earlier implementations were restricted to amplitude-modulated fields. Second, previous schemes were bandwidth-limited by the optical chirp rate. In our case, the limit is set by the inhomogeneous width $\Gamma_{inh}$ and by the spectral span of the second and third pulses, a far less restrictive condition. Yet, our approach shares the same limitation with earlier protocols regarding signal duration: the first and second fields must illuminate the atomic medium within the optical coherence lifetime $T_2$ to record spectral interferences, limiting both signal and control pulse lengths, and the echo field envelope decays as $e^{\frac{-2t}{T_2}}$. Finally, the analytical echo field expression (Equation~\ref{eq:the_equation}) holds only in the weak field limit. In this regime, the photon echo remains weak but can be efficiently detected via heterodyne methods~\cite{beaudoux2011}.

\section{Experimental demonstration}
\subsection{Experimental setup}
\begin{figure}[t]
\includegraphics[width=8.cm]{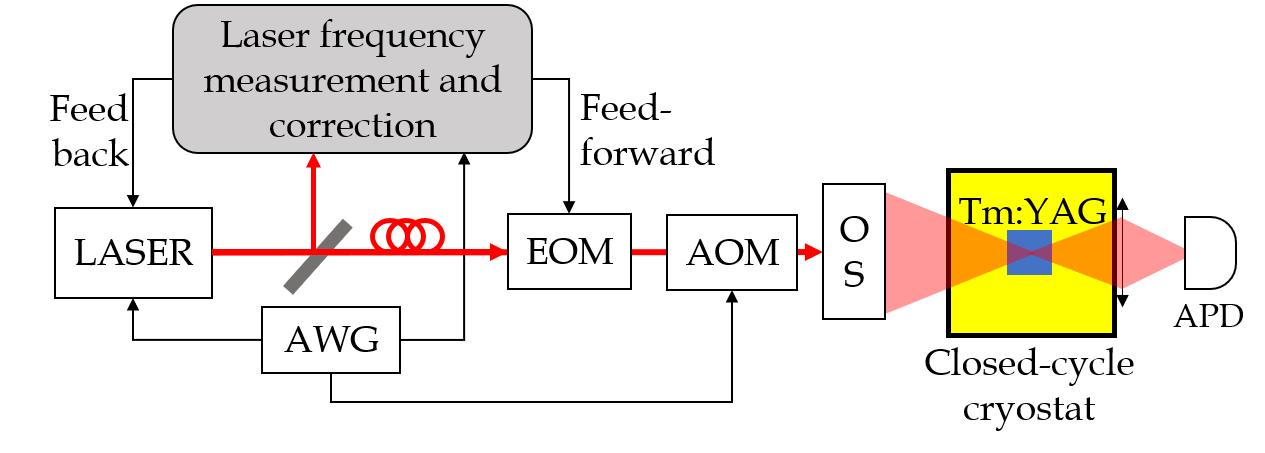}
\caption{Experimental setup. EOM: electro-optical modulator. AOM: acousto-optic modulator. OS: optical system. APD: avalanche photodetector. AWG: arbitrary waveform generator. }
\label{fig:setup}
\end{figure}

The experimental setup is shown in Figure~\ref{fig:setup}. The laser is a cw Distributed Bragg Reflector source (Photodigm DBR795PH), externally modulated to perform frequency excursions. These frequency excursions are controlled with a versatile, four-stage correction method including pre-distorsion of the voltage command, feedback and feed-forward real-time correction~\cite{llauze2024versatile}. This allows us to routinely obtain MHz-scale errors on steep excursions and laser linewidth reduction down to $30$~kHz during fixed frequency operation. The DBR chip is cooled down to $5^\circ$C in order to reach the absorption wavelength of Tm$^{3+}$ ions in YAG ($793.379$ nm in vacuum).

A $3\times4\times5$mm$^3$, $0.5\% at.$Tm:YAG single crystal (FLIR Scientific Materials) is placed in the vacuum chamber of a closed-cycle, low-vibration cryocooler (Montana Instruments s50 Cryostation). Due to the coupling of the optical lines to mechanical strain~\cite{louchet2019piezospectroscopic}, further vibration damping is required and a home-made vibration-damping platform is used. In such conditions, the crystal reaches a $3.5$~K temperature and exhibits an inhomogeneous broadening $\Gamma_{inh}=17$~GHz, a peak absorption $\alpha L=2$ and an optical coherence lifetime $T_2\simeq 10~\mu$s. The crystal is illuminated by a single laser beam, temporally shaped using a double-pass acousto-optic modulator (AOM, AA Opto-Electronic). The laser beam is focused on a $w_0=45\pm 5~\mu$m radius spot in the Tm:YAG crystal, using an optical system composed of a telescope (beam expander) and a lens. The maximum power on the crystal is about $5$~mW.

\begin{figure}[t]
\includegraphics[width=8.cm]{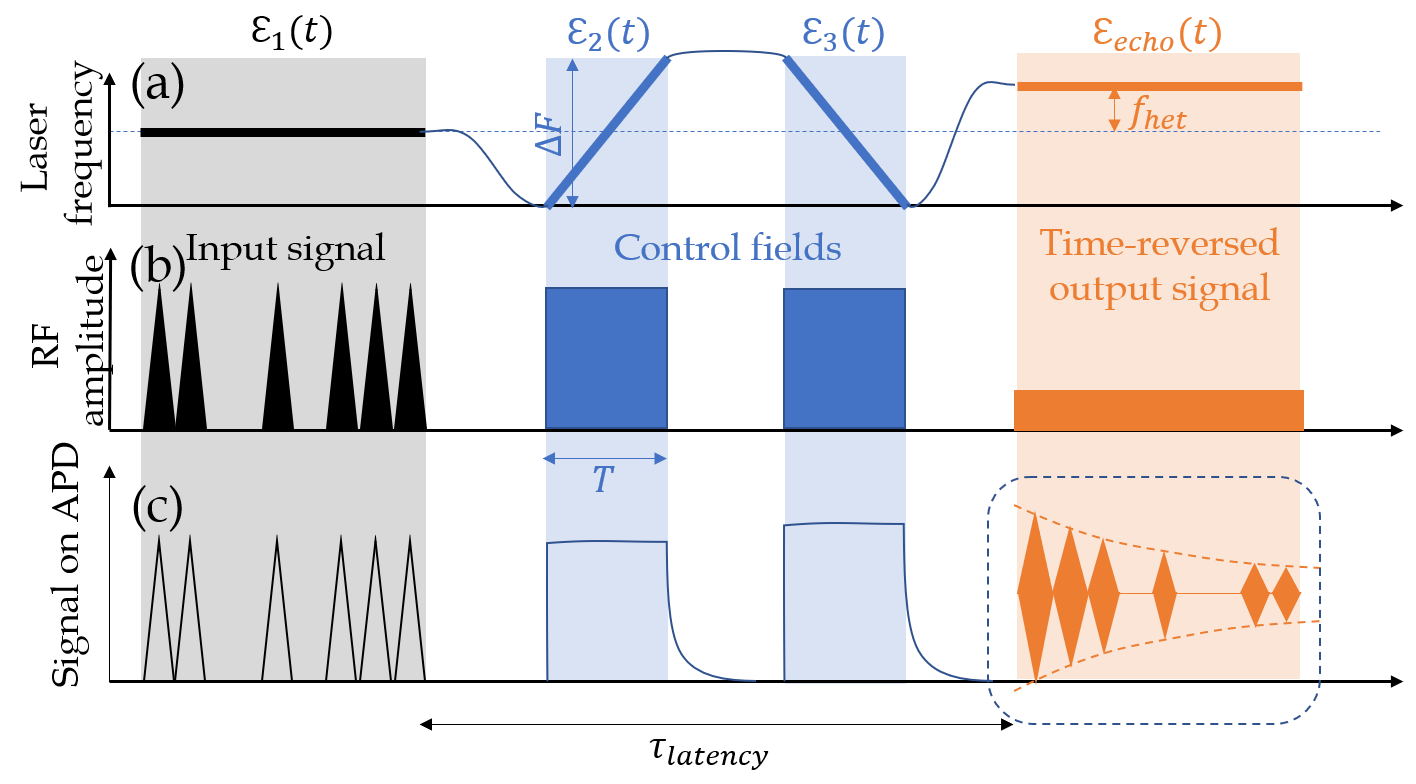}
\caption{Schematic pulse sequence leading to the spontaneous, coherent emission of a phase-preserving time-reversed copy of the input field $\mathcal{E}_1(t)$. (a) Laser instantaneous frequency. (b) Amplitude of the RF signal sent to the AOM, enabling temporal shaping of the cw laser. (c) Signal detected after transmission through the crystal. The control pulses are followed by a free-induction decay radiation~\cite{allen1987}. A heterodyne pulse makes the time-reversed signal appear as a phase-stable beatnote. The latency time $\tau_{latency}$ is defined as the interval between the end of the input signal and the beginning of the output signal.  }
\label{fig:timesequence}
\end{figure}

The pulse sequence is shown in Figure~\ref{fig:timesequence}. During the first part of the excitation sequence, the laser frequency is constant ($\nu_{laser}=\nu_0$) and the optical field is amplitude-, phase- or frequency-modulated using the AOM. Due to the double-pass configuration, the resulting dephasing and frequency shift of the optical wave is doubled. In the second part, the first control pulse is generated by linearly chirping the laser emission over an interval $\Delta F$ around $\nu_0$. Typically, the duration of this chirp is $T=5~\mu$s and the chirp span is $\Delta F=40$~MHz. Then, after a $30~\mu$s waiting time, the second control pulse is generated, composed of a linear laser chirp with the opposite chirp rate. The echo carrying the time-reversed signal is then spontaneously emitted by the atomic ensemble. In the temporal window where the echo is expected, a rectangular pulse is generated with the AOM, while the laser is tuned to $\nu_0+30~$MHz, enabling the heterodyne detection of the echo on a Thorlabs APD110A avalanche photodiode. The resulting beatnote enables the restoration of the signal into the RF domain. As the heterodyne pulse is sent in the same spatial mode as the echo emission, the beatnote is inherently stable, at the price of some perturbation of the atomic coherences. The time sequence is entirely controlled by an arbitrary waveform generator (Tektronix AWG5004).

\subsection{Phase-shift-keying}
We test our scheme using an input signal composed of 17 gaussian pulses with successive $\pi/4$ phase jumps. 
The output signal reproduces the succession of pulses but with a visible envelope decay, compatible with the optical coherence relaxation. The phase of each pulse should reproduce the input phase in a time-reversed order and with a sign inversion due to the complex conjugate in Eq.~\ref{eq:E1}. We compute the phase of each pulse by digitally demodulating the beatnote signal, and verify that the output phase variations closely follow the expected behaviour.
Some dispersion of the detected phase is observed, especially at long delays, appearing as the buildup of a temporally linear phase offset. We ascribe this to a frequency instability in the AWG or oscilloscope internal clock.

\begin{figure}[t]
\includegraphics[width=8.cm]{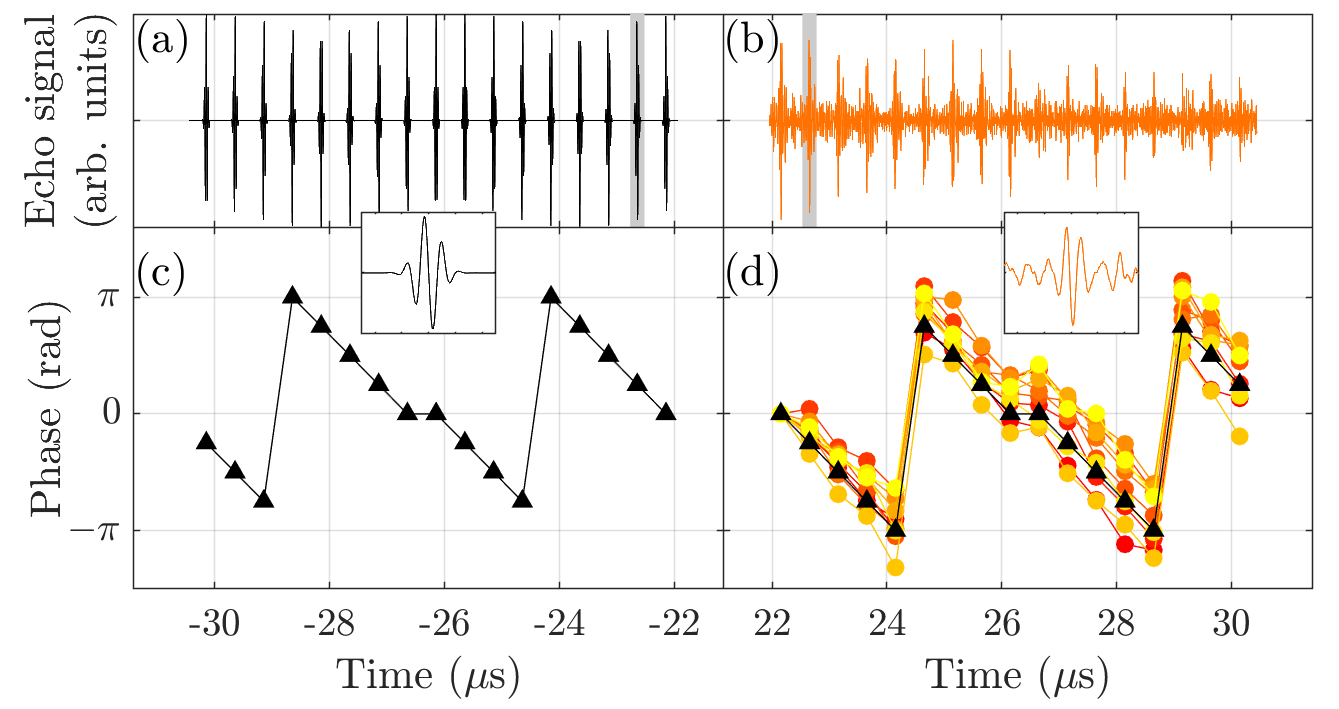}
\caption{Time-reversal of a 8-$\mu$s-long PSK-encoded pulse sequence, where each pulse has a $50$~ns FWHM duration. (a) Input signal. (b)  Example of experimental output signal obtained by heterodyne detection. (c) Phase of each pulse of the input signal. (d) Measured phase of the output signal, obtained by digital demodulation of each pulse (color circles). The experiment is conducted 10 times to assess its repeatability. The phases are referenced to the first pulse of the time-reversed signal and compared to the expected values, shown as black triangles. Insets: close-up on one of the pulses, highlighted by the gray rectangle in panels (a) and (b). }
\label{fig:psk}
\end{figure}

\subsection{Frequency-shift-keying}
\begin{figure}[t]
\includegraphics[width=8.cm]{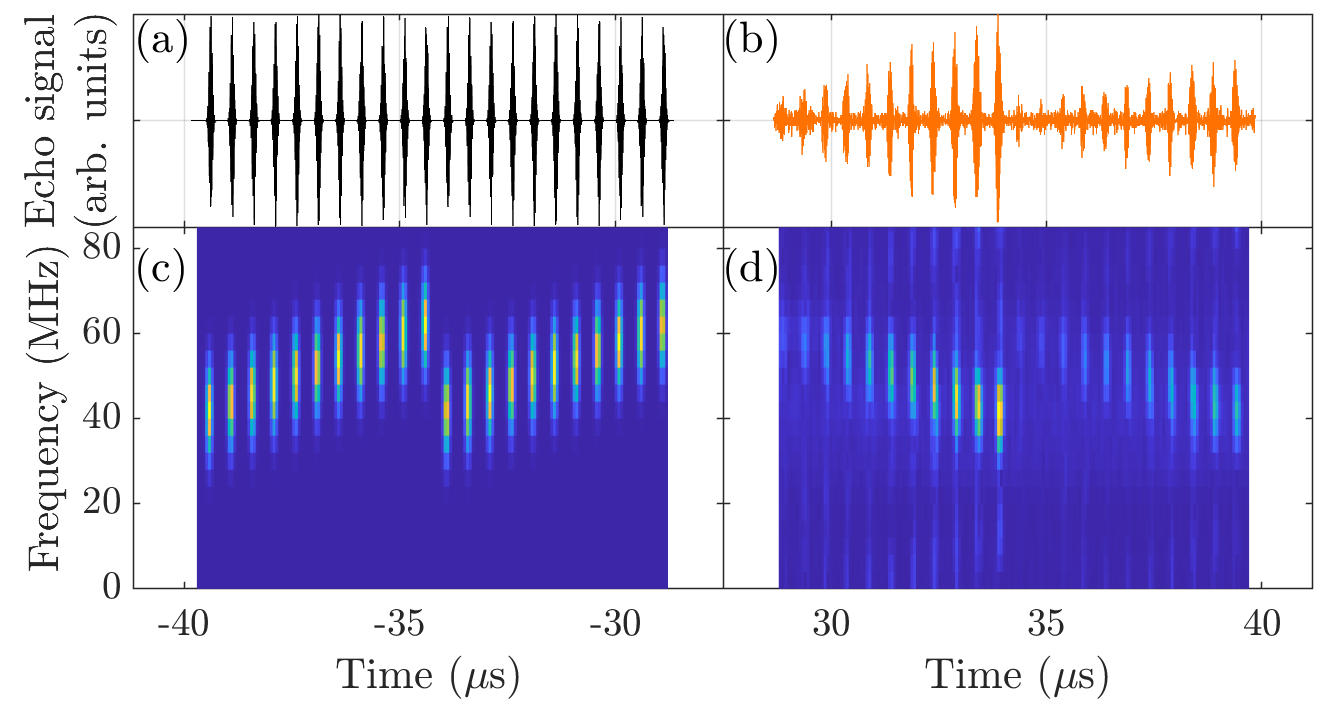}
\caption{Time-reversal of a 12-$\mu$s-long,  FSK-encoded 22-pulse input signal. (a) Input signal. (b) Example of experimental output signal obtained by heterodyne detection. The output beatnote is affected by the $50$~MHz cutoff of the photodetector. The spectrograms are calculated on the input signal (c) and on the experimental signal, averaged over 10 repetitions (d).}
\label{fig:fsk}
\end{figure}

We also test our time-reversal scheme using an input signal composed of 22 gaussian pulses with frequencies stepped by $2$~MHz from $20$ to $0$~MHz (see Figure~\ref{fig:fsk}). A comparison of the spectrograms confirms that the output reproduces the frequency steps. The demonstrated bandwidth largely exceeds the $\sqrt{r}=\sqrt{\Delta F/T}=2.8$~MHz limitation that would have been observed with previous photon echo-based approaches~\cite{mossberg1982,wang1995optical,linget2013}. The $50$~MHz cutoff  is caused by the photodetector, effectively limiting the time-reversal bandwidth in this demonstration.

\subsection{Towards broadband time-reversal}

To time-reverse broadband signals, the control fields must span a broader spectral range. Increasing the chirps span $\Delta F$ while keeping their duration $T$ constant reduces the density of energy deposited by the control fields on each spectral class of ions, leading to a $1/\Delta F$ attenuation of the echo field. This loss can be mitigated by increasing $T$, provided it remains shorter than $T_2$.
However, increasing the total energy deposited in the sample eventually reduces the echo field amplitude due to instantaneous spectral diffusion (ISD)~\cite{liu1990}. More specifically, the homogeneous linewidth is increased proportionally to the volumic density of ions who were promoted to a different state. We use the formalism developed in~\cite{thiel2014measuring} to express the effective homogeneous linewidth: $\Gamma_{h,eff}=\Gamma_h + \frac{\beta_{ISD}}{2} |\Delta n_e|$,
where $\beta_{ISD}$ is a material-dependent coefficient and $|\Delta n_e|$ is the volumic density of ions that have been promoted to a different electronic state. In the case of two fields of duration $T$ addressing a spectral interval $\Delta F$ at the center of an inhomogeneously broadened absorption profile, this expression can be rewritten as:
\begin{equation}
\Gamma_{h,eff}= \frac{1}{\pi T_2} + \Gamma_{ISD}(1-e^{-c_T T })
\label{eq:Gammaheff}
\end{equation}
where
\begin{equation}
\Gamma_{ISD}= \frac{\beta_{ISD}n_{ions}}{\pi} \frac{ \Delta F}{\Gamma_{inh}} \textrm{ and }
c_T       = \frac{2\pi \alpha \phi}{ n_{ions}}\frac{\Gamma_{inh}}{\Delta F}
\end{equation}
with $\alpha$ the absorption coefficient and $\phi$ the photon surface flux. In Tm:YAG, $\beta_{ISD}=2.4\times10^{-12}$~Hz.cm$^3$~\cite{thiel2014measuring}. With a $0.5\%$ Tm concentration, $\alpha=400$~m$^{-1}$, and $n_{ions}=6.9\times10^{25}$~m$^{-3}$. With $\phi=3.2\times 10^{24}$~m$^{-2}$.s$^{-1}$ and $\Delta F=40$~MHz this yields $\Gamma_{ISD}=124$~kHz and $c_T=60.7$~kHz.

\begin{figure}[t]
\includegraphics[width=8.cm]{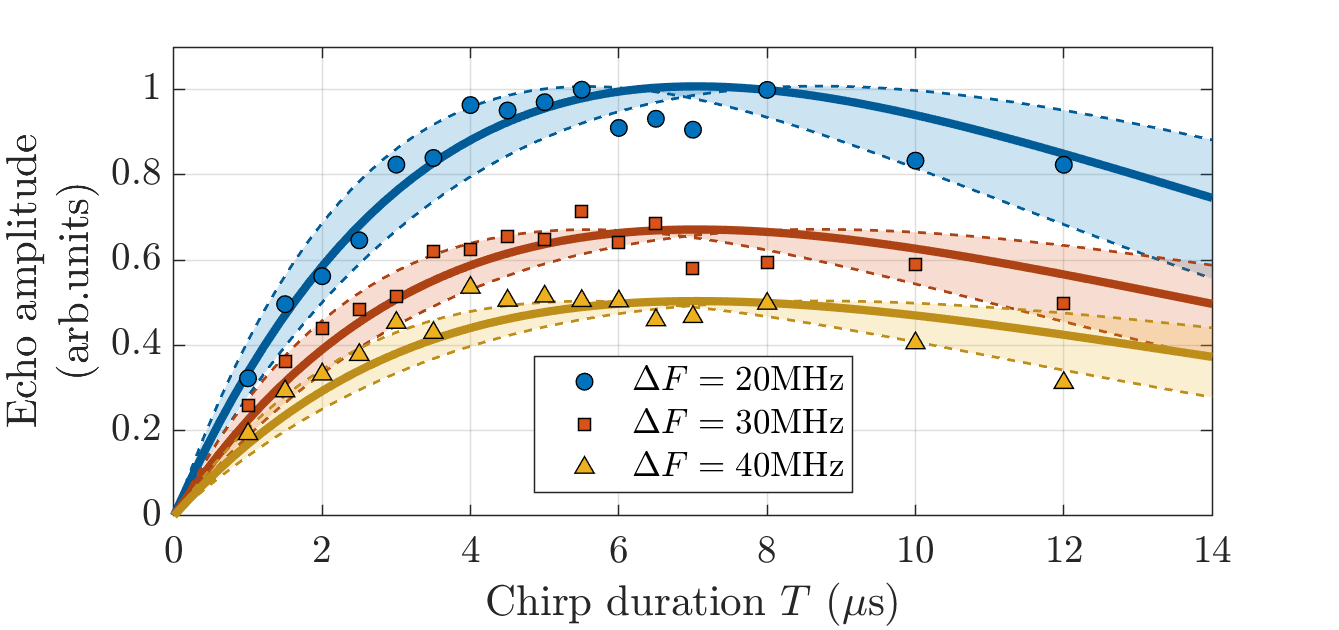}
\caption{Effect of the chirp span and duration on the echo amplitude revealing the presence of instantaneous spectral diffusion. Symbols: amplitude of the measured echo envelope. Lines: theoretical behaviour given by Equation~\ref{eq:Eecho} (see text for details), with a beam waist $w_0=45~\mu$m (solid line), adjusted to the experimental data using a unique vertical scaling factor for each waist value. The uncertainty on the determination of the waist $w_0$ is reflected as dashed lines. 
}
\label{fig:largebw}
\end{figure}

In the limit of weak pulses ($c_T T\ll 1$), we simplify this expression: $1-e^{-c_T T}\simeq c_T T$.
The photon echo field amplitude is directly affected by the effective homogeneous linewidth:
\begin{equation}
|\mathcal{E}_{echo}| \propto \frac{T}{\Delta F} e^{2\pi  t_{12} \Gamma_{h,eff} } \simeq \frac{T}{\Delta F} e^{-2\pi t_{12} \Gamma_{ISD} c_T T }
\label{eq:Eecho}
\end{equation}
where $t_{12}$ is the time between the input signal and the beginning of the control field $\mathcal{E}_2(t)$.
To verify this, we use our architecture to time-reverse a signal composed of a single gaussian pulse sent $t_{12}=3~\mu$s before the first control field and study the evolution of the echo amplitude when the chirp span $\Delta F$ and the chirp duration $T$ are varied (Figure~\ref{fig:largebw}). We find that the results align well with the model (Equation~\ref{eq:Eecho}): the echo linearly grows with $T$, then reaches a maximum around $T\simeq5~\mu$s and then decays. This confirms the significant contribution of ISD to the effective homogeneous linewidth, and the corresponding difficulty to reach large bandwidths for the time-reversal scheme.


\section{Discussion}
Through these examples we have shown that our fully analog, photon-echo-inspired architecture enables the faithful retrieval of the time-reversed radio-frequency field. Aside from a gradual output signal envelope decay due to decoherence, the input signal's amplitude and phase signatures are well recovered.
In addition, the time-reversal scheme is compatible with $10~\mu$s-long signals and operates with latency times of a few tens of $\mu$s. Note that the duty cycle of our time-reversal processor is inherently limited ($\sim0.1\%$) by the long atomic population decay time.

While this time-reversal scheme is promising, the need for a cryogenic apparatus may appear as a strong deterrent for applications, especially when dealing with multiple receivers. Nevertheless, the four-wave mixing nature of the photon echo process enables simultaneous processing of multiple signals using an angular mutiplexing approach, allowing for inherent scalability with minimal added complexity or power consumption~\cite{barber2017spatial}. 

The ISD-based limitation on the time-reversal processing bandwidth can be mitigated by reducing the dopant concentration. However, compensating for the resulting loss in optical depth requires longer samples, which, in turn, limits the available Rabi frequency. In that context, guided photonic platforms offer a promising alternative to bulk crystals, potentially enabling zero beam divergence and enhanced light-matter interaction over several mm~\cite{zhou2023photonic}. Notably, the pursuit of efficient photon echoes over large bandwidths aligns with ongoing efforts in developing broadband quantum memories~\cite{dajczgewand2014,ortu2022multimode}.

\section{Conclusion}
In this work, we have demonstrated a successful proof-of-principle of low-latency, fully analog time-reversal of radiofrequency fields. This process is based on a modified version of three-pulse photon echo and leverages the spectroscopic properties of a cryogenically-cooled rare-earth ion-doped crystal.

Although this demonstration operates within a limited bandwidth, constrained  technically by the detection and fundamentally by instantaneous spectral diffusion, it validates several key features, namely the ability to time-reverse frequency- and phase-modulated signals, the compatibility with $\mu$s-scale signals, and a latency less than $100~\mu$s. The phase-preserving property validates this work as the first analog approach that could be practically applied to wave refocusing in complex media, in contrast to previous analog time-reversal schemes which were only able to handle amplitude-modulated signals~\cite{coppinger1999,zhang2015,linget2013}. Combined with the low achieved latency, this work opens new perspectives for real-time focusing of radiofrequency waves in dynamic complex media.


\section{Back matter}

\begin{backmatter}
\bmsection{Funding} (automatic) The authors acknowledge support from the French National Research Agency (ANR)
through the projects ATRAP (ANR-19-CE24-0008), CHORIZO (ANR-24-CE47-1190), from Region Île-de-France in the framework of DIM QuanTiP, and from the program “Investissements d’Avenir” launched by the French Government.

\bmsection{Acknowledgment} The authors are grateful to Félix Montjovet-Basset for his contribution at early stages of the project, and to Daniel Civiale for technical support, and acknowledge helpful discussions with Jean-Pierre Huignard and Jean-Louis Le Gouët.

\bmsection{Disclosures} The authors declare no conflicts of interest.

\bmsection{Data availability} Data underlying the results presented in this paper are not publicly available at this time but may be obtained from the authors upon reasonable request.

\end{backmatter}



\bibliography{toute_la_biblio}

\bibliographyfullrefs{toute_la_biblio}


\ifthenelse{\equal{\journalref}{aop}}{%
\section*{Author Biographies}
\begingroup
\setlength\intextsep{0pt}
\begin{minipage}[t][6.3cm][t]{1.0\textwidth} 
  \begin{wrapfigure}{L}{0.25\textwidth}
    \includegraphics[width=0.25\textwidth]{john_smith.eps}
  \end{wrapfigure}
  \noindent
  {\bfseries John Smith} received his BSc (Mathematics) in 2000 from The University of Maryland. His research interests include lasers and optics.
\end{minipage}
\begin{minipage}{1.0\textwidth}
  \begin{wrapfigure}{L}{0.25\textwidth}
    \includegraphics[width=0.25\textwidth]{alice_smith.eps}
  \end{wrapfigure}
  \noindent
  {\bfseries Alice Smith} also received her BSc (Mathematics) in 2000 from The University of Maryland. Her research interests also include lasers and optics.
\end{minipage}
\endgroup
}{}

\end{document}